\documentclass{aastex63}
\received{}
\revised{}
\accepted{}
\submitjournal{ApJ}

\usepackage{amsmath}

\shorttitle{High-Resolution Spectroscopy of N132D}
\shortauthors{Suzuki et al.}
\begin{document}

\title{
Plasma Diagnostics of the Supernova Remnant N132D Using Deep 
XMM--Newton Observations with the Reflection Grating Spectrometer 
}

\correspondingauthor{Hitomi Suzuki}
\email{suzuki-hitomi@ed.tmu.ac.jp}

\author{Hitomi Suzuki}
\affiliation{Department of Physics, Tokyo Metropolitan University, 
1-1 Minami-Osawa, Hachioji, Tokyo 192-0397, Japan}
\affiliation{The Institute of Space and Astronautical Science (ISAS), Japan Aerospace and Exploration Agency (JAXA),
3-1-1 Yoshinodai, Chuo-ku, Sagamihara 299-8510, Japan}

\author{Hiroya Yamaguchi}
\affiliation{The Institute of Space and Astronautical Science (ISAS), Japan Aerospace and Exploration Agency (JAXA),
3-1-1 Yoshinodai, Chuo-ku, Sagamihara 299-8510, Japan}
\affiliation{Department of Physics, The University of Tokyo,
7-3-1 Hongo, Bunkyo-ku, Tokyo 113-0033, Japan}

\author{Manabu Ishida}
\affiliation{Department of Physics, Tokyo Metropolitan University, 
1-1 Minami-Osawa, Hachioji, Tokyo 192-0397, Japan}
\affiliation{The Institute of Space and Astronautical Science (ISAS), Japan Aerospace and Exploration Agency (JAXA),
3-1-1 Yoshinodai, Chuo-ku, Sagamihara 299-8510, Japan}

\author{Hiroyuki Uchida}
\affiliation{Department of Physics, Kyoto University, 
Kitashirakawa Oiwake-cho, Sakyo-ku, Kyoto 606-8502, Japan}

\author{Paul P.\ Plucinsky}
\affiliation{Center for Astrophysics | Harvard \& Smithsonian, 60 Garden St., Cambridge, MA 02138, USA}

\author{Adam R.\ Foster}
\affiliation{Center for Astrophysics | Harvard \& Smithsonian, 60 Garden St., Cambridge, MA 02138, USA}

\author{Eric D.\ Miller}
\affiliation{Kavli Institute for Astrophysics and Space Research, Massachusetts Institute of Technology, 77 Massachusetts Ave., Cambridge, MA 02139, USA}


\begin{abstract}
We present XMM--Newton observations of N132D, the X-ray brightest supernova remnant (SNR) 
in the Large Magellanic Cloud (LMC), using the Reflection Grating Spectrometer (RGS) 
that enables high-resolution spectroscopy in the soft X-ray band. 
A dozen emission lines from L-shell transitions of various elements at intermediate charge 
states are newly detected in the RGS data integrating the $\sim$200-ks on-axis observations.
The 0.3--2.0-keV spectra require at least three components of thermal plasmas with 
different electron temperatures and indicate clear evidence of non-equilibrium ionization (NEI). 
Our detailed spectral diagnostics further reveal that the forbidden-to-resonance line ratios 
of O\,{\footnotesize VII} and Ne\,{\footnotesize IX} are both higher than expected for 
typical NEI plasmas. 
%%追加20200707 -> Done. Either は二者択一にしか使えないことに注意。
This enhancement could be attributed to either resonance scattering or
emission induced by charge exchange in addition to a possible
contribution from the superposition of multiple temperature components,
although the lack of spatial information
prevents us from concluding which is most likely.%%
\end{abstract}



\section{Introduction} \label{sec:introduction}

High-resolution spectroscopy of supernova remnants (SNRs) using X-ray grating spectrometers 
has enabled detailed plasma diagnostics, providing insights into the shock physics and radiation 
processes in the hot plasmas that consist of the supernova ejecta and ambient medium
\citep[e.g.,][]{Vink2003,Katsuda2012,Miceli2019,Uchida2019}. 
In particular, the Reflection Grating Spectrometer (RGS) on board the XMM--Newton satellite 
is suitable for observing moderately extended objects (a few arcmin in diameter), 
like SNRs in the Large Magellanic Cloud (LMC) because of its relatively large dispersion angle 
\citep{Herder2001,Rasmussen2001}. 
N132D, a middle-aged SNR located in the stellar bar of the LMC, is one of such objects, owing to 
its high luminosity in the X-ray band as well as the moderate angular diameter of $\sim 2'$. 
In this paper, we present high-resolution spectroscopy of this SNR, utilizing the excellent features 
of the RGS.

Optical studies of N132D estimated its age to be 2500--3100\,yrs based on the expansion of 
the O-rich knots \citep{Morse1995,Vogt2011,Law2020}. 
Hubble Space Telescope observations revealed 
the strong emission of the C/Ne-burning products (i.e., O, Ne, Mg) with little emission of 
the O-burning products (i.e., Si, S), leading to an interpretation of a Type Ib supernova origin 
for this SNR \citep{Blair2000}. 
The X-ray emission from N132D was first identified by the Einstein Observatory, 
which also revealed the clear shell-like morphology \citep{Mathewson1983}. 
Using ASCA, \cite{Hughes1998} found that the elemental abundances of the entire SNR 
are consistent with the mean LMC values \citep{Russell1992}, suggesting that the emission 
in the soft X-ray band ($\lesssim$\,2\,keV) is dominated by the swept-up interstellar medium (ISM). 
On the other hand, the Fe K emission, first detected by BeppoSAX \citep{Favata1997}, 
shows a more centrally concentrated morphology \citep{Behar2001,Borkowski2007,Sharda2020}, 
suggesting an ejecta origin of this emission. 
The Suzaku and NuSTAR studies of this SNR detected Ly$\alpha$ emission from 
H-like Fe, alongside the stronger He-like K-shell emission \citep{Bamba2018}. 
Notably, the K-shell emission from highly-ionized Fe is a common feature of core-collapse SNRs 
\citep{Yamaguchi14}, consistent with the optical studies of this SNR \citep[e.g.,][]{Vogt2011}. 
More recently, Hitomi observations utilizing the remarkably high spectral resolution of the 
micro-calorimeter (but with an extremely short exposure) revealed that the Fe K emission 
is substantially redshifted with respect to the local LMC ISM \citep{2018PASJ7016H}, 
implying an asymmetric explosion of the progenitor of this SNR.

Similarly to other core-collapse SNRs, N132D is located within a dense environment. 
Infrared and radio studies revealed that the south rim of this SNR is interacting with 
particularly dense clouds \citep{Williams2006,Sano2015}, 
whose pre-shock density is estimated to be 100--500\,cm$^{-3}$, 
depending on the direction \citep{Dopita2018}. 
X-ray studies also indicated that the SNR has expanded into a cavity formed by 
the pre-explosion stellar wind activity of the progenitor and reached the cavity wall 
recently \citep{Hughes1998,Chen2003}. 
The strong interaction between the SNR shock and dense clouds is also suggested by 
the bright GeV and TeV $\gamma$-ray emission possibly originating from hadronic processes
\citep{Ackermann2016,HESS2015}; N132D is indeed the brightest in $\gamma$-ray among 
all known SNRs \citep{Acero2016}. We may, therefore, independently find a signature of such 
SNR-ISM interaction in its soft X-ray spectrum, which is one of our motivations for the present study.

The early RGS observations of N132D, conducted as a part of the Performance Verification 
program of XMM--Newton with the total effective exposure of $\sim$\,50\,ks, were reported by 
\cite{Behar2001}. They identified the spectral lines of C, N, O, Ne, Mg, Si, S, and Fe in 6--37\,\AA\ 
(corresponding to the photon energy of 0.34--2.0\,keV) and revealed the wide range of Fe charge 
states (Fe$^{16+}$--Fe$^{21+}$) that indicates the presence of multiple electron temperatures. 
In the previous work, however, first- and second-order spectra with different resolutions were summed in the analysis, 
and no quantitative spectral modeling was performed to determine whether the plasma was in collisional ionization equilibrium 
(CIE) or non-equilibrium ionization (NEI). Moreover, while N132D has been routinely observed by 
XMM--Newton for the instrumental calibration purpose, the scientific outcome from these 
observations (whose net exposure exceeds 1\,Ms) has never been reported to date. 
Thus, here we analyze all the available RGS data with a more quantitative approach 
that simultaneously models first and second order.
In particular, the high-resolution second-order spectrum enables detailed diagnostics using 
the well-separated emission lines, revealing clear evidence of NEI and a possible 
signature of additional effects that have modified the line intensity ratios between the forbidden 
and resonance emission, such as resonance scattering and/or charge exchange.

This paper is organized as follows. 
Sections 2 and 3 describe the data reduction and response generation, respectively. 
In Section 4, we perform spectral analysis based on realistic plasma models. 
We then conduct detailed spectral diagnostics using several strong lines and 
discuss the results in Section 5. Finally, we conclude this study in Section 6. 
Throughout the paper, we assume the distance to the LMC to be 50\,kpc
\citep{Pietrzynski2013}. 
The errors quoted in the text and table and error bars given in the figures 
represent the 1$\sigma$ confidence level, unless otherwise stated.

\section{Data Reduction} \label{sec:observation}
The primary purposes of the routine observations of N132D with XMM-Newton are the gain, spectral redistribution, and effective area calibrations of the European Photon Imaging Camera (EPIC). 
Therefore, although the high-resolution spectral data with the RGS were simultaneously obtained, 
the SNR was placed outside or near the edge of the field of view of the RGS in most of these observations. 
Since our aim is to analyze the high-resolution grating spectra, we use only the datasets 
where the target was observed on the optical axis of the telescope. 

The data are reprocessed using the {\tt rgsproc} task in the XMM--Newton Science Analysis Software 
(XMMSAS) version 18.0.0 and the Current Calibration Files (CCF). We extract background light curves 
using the standard procedure\footnote{https://xmm-tools.cosmos.esa.int/external/xmm\_user\_support/documentation/sas\_usg/USG/rgslightcurve.html} 
and filter out observation periods when the background count rate is higher than 0.1\,cnt\,s$^{-1}$. 
In addition, we discard Observation IDs whose exposure time after the screening is less than 10\,ks.
The observations that meet these criteria are given in Table \ref{tab:datainfo}, 
obtaining the total effective exposure of 193\,ks for both the RGS1 and RGS2. 

\begin{figure*}
\gridline{\fig{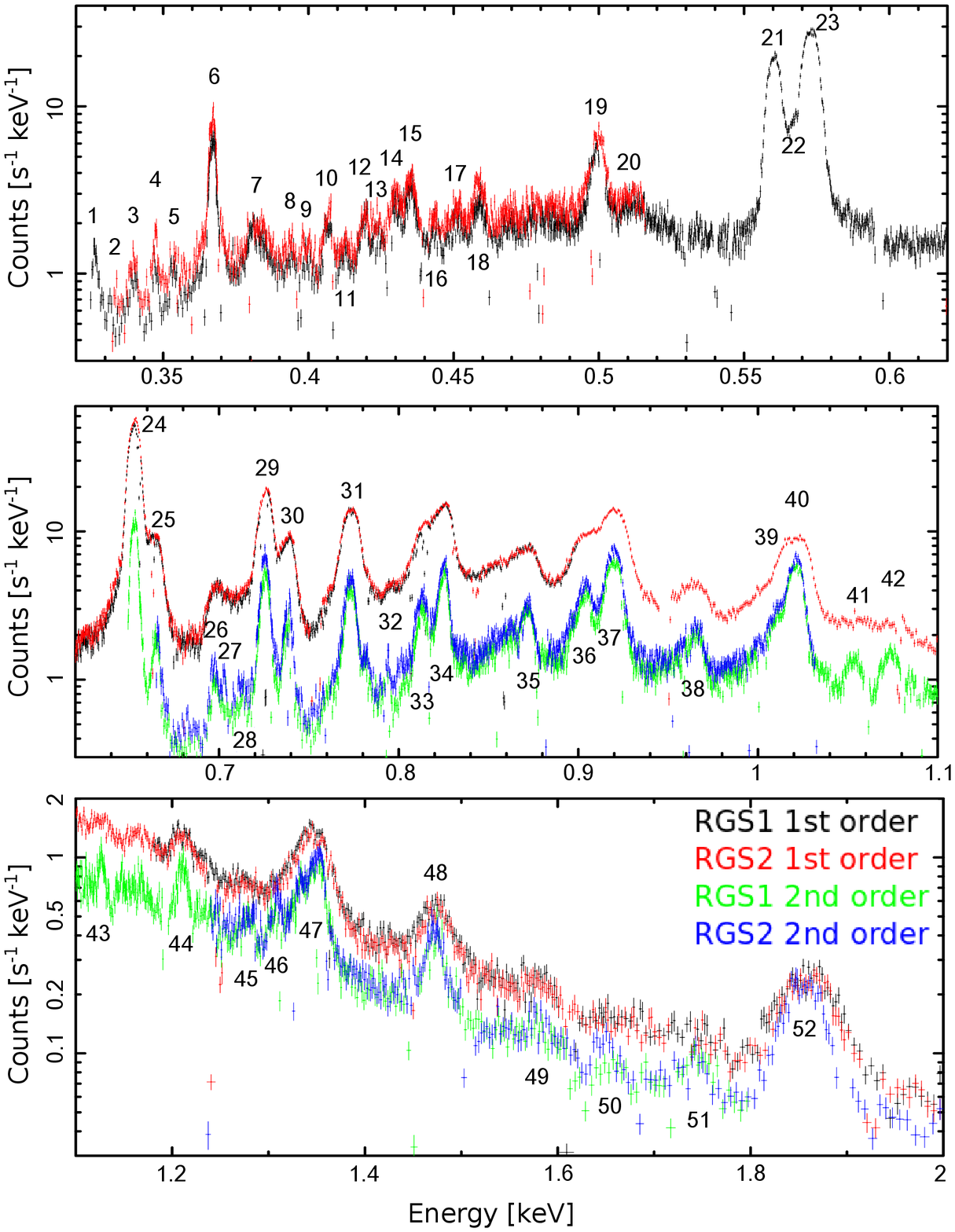}{0.75\textwidth}{}}
\caption{XMM--Newton RGS spectra where all the observations given in Table\,\ref{tab:datainfo} are integrated.  
Black and red are the first-order spectra of the RGS1 and RGS2, respectively. 
Green and blue are the second-order spectra of these spectrometers, 
whose bandpass is limited to $\gtrsim 0.65$\,keV. 
The detected lines are labeled and listed in Table~\ref{tab:spectrum_9data_lines} with their identification.
\label{fig:spectrum_9data}}
\end{figure*}

We extract the first- and second-order RGS spectra of each observation from the region centered on
(RA, Dec)$_{\rm J2000}$ = (05 25 03.96, --69 38 29.6) with the cross-dispersion width of $2.\!'2$ 
to contain the entire SNR regardless of the roll angle. 
Figure \ref{fig:spectrum_9data} shows the resulting spectra, 
where the data from all the observations (193\,ks) are integrated to improve the photon statistics. 
The detected emission lines are summarized in Table \ref{tab:spectrum_9data_lines},  
together with their centroid energy and identified atomic transition. 
In addition to the lines already reported by \cite{Behar2001}, we have newly detected 
a dozen L-shell emission from Fe and the intermediate-mass elements as well as 
weak K-shell emission of high-$n$ transitions. 
Notably, the Ne\,{\footnotesize IX} forbidden and resonance lines are clearly resolved 
for the first time, owing to the better energy resolution of the second-order spectrum 
($\Delta$E = 9.9\,eV at 0.9\,keV for an extended object with an angular diameter of $2.\!'2$, 
which is about twice better than the first-order's energy resolution).
This allows us to put strong constraints on the important spectral parameters, such as 
electron temperature and ionization timescale, as demonstrated in the following sections.

\begin{deluxetable*}{llcc}
\tablecaption{Observation log. \label{tab:datainfo} }
\tablewidth{0pt}
\tablehead{
\colhead{Observation ID} &\colhead{Date} & \colhead{Roll angle} & \colhead{Exposure$^{\ast}$} \\
\colhead{ } &\colhead{ } & \colhead{(deg)} & \colhead{(ks)} }
\startdata
      0125100201&2000 May 23& 339.77 &13.63 \\
      0157160601&2002 Nov 16& 152.29& 26.20\\
      0157160801&2002 Nov 24& 160.35& 29.09\\
      0157161001& 2002 Dec 14& 180.55& 29.91\\
      0157360201&2002 Dec 31& 196.82& 15.02\\
      0157360301&2003 Jan 17& 215.16& 28.66\\
      0157360501&2003 Feb 22& 251.66& 14.92\\
      0129341301&2003 Sep 13& 88.84& 20.34\\
      0137551101&2004 Sep 8& 88.76& 15.04\\ \hline
\enddata
\tablecomments{$^{\ast}$Effective exposure after the screening (see text).}
\end{deluxetable*}

\begin{deluxetable*}{lllllllllllllllll}
\tablecaption{Detected emission lines and their identification. \label{tab:spectrum_9data_lines}}
\tablehead{
\colhead{Label} & \colhead{Energy} & \colhead{Ion}& \colhead{Transition}& \colhead{   }&\colhead{Label} &  \colhead{Energy} & \colhead{Ion} & \colhead{Transition}& \colhead{   }&\colhead{Label} & \colhead{Energy} & \colhead{Ion}& \colhead{Transition}& \colhead{   }\\
\colhead{ } & \colhead{(keV)} & \colhead{ } & \colhead{ }& \colhead{   }&\colhead{ } & \colhead{(keV)} & \colhead{ }& & \colhead{   }& \colhead{   }&\colhead{(keV)} & \colhead{ } & \colhead{ }& \colhead{   }}
\startdata
	1 $^{\ast}$& 0.329  & S\,{\footnotesize XIII }& 3s $\rightarrow$ 2p&&		20 $^{\ast}$& 0.511&S\,{\footnotesize XIV }&4d $\rightarrow$ 2p&&			37 & 0.917&Fe\,{\footnotesize XIX }&3d $\rightarrow$ 2p\\
	2 $^{\ast}$& 0.332  & Si\,{\footnotesize XI }& 4d $\rightarrow$ 2p&&		21 & 0.561&O\,{\footnotesize VII } &2s $\rightarrow$ 1s (f)&&				     & 0.922&Ne\,{\footnotesize IX }&2p $\rightarrow$ 1s (r)\\				
	3 $^{\ast}$& 0.341 & S\,{\footnotesize XII }&3d $\rightarrow$ 2p&&		22 & 0.569&O\,{\footnotesize VII } &2p $\rightarrow$ 1s (i)&&				38 & 0.964&Fe\,{\footnotesize XX }&3d $\rightarrow$ 2p\\	
	4 $^{\ast}$& 0.347  & Ca\,{\footnotesize XI }&3s $\rightarrow$ 2p&&		23 & 0.574&O\,{\footnotesize VII } &2p $\rightarrow$ 1s (r)&&				     & 0.965&Fe\,{\footnotesize XX }&3d $\rightarrow$ 2p\\		
	5 $^{\ast}$& 0.352  & Ca\,{\footnotesize XI }&3s $\rightarrow$ 2p&& 		24 & 0.654&O\,{\footnotesize VIII } &2p $\rightarrow$ 1s&&			   	39 $^{\ast}$& 1.009&Fe\,{\footnotesize XXI }& 3d $\rightarrow$ 2p \\
	6 & 0.368 & C\,{\footnotesize VI } &2p $\rightarrow$ 1s&&				25 & 0.666&O\,{\footnotesize VII } &3p $\rightarrow$ 1s&&				40 & 1.022&Ne\,{\footnotesize X }&2p $\rightarrow$ 1s\\						
	7 & 0.381& S\,{\footnotesize XIV }&3d $\rightarrow$ 2p&&				26 $^{\ast}$&0.698&O\,{\footnotesize VII }& 4p $\rightarrow$ 1s &&			41 & 1.053&Fe\,{\footnotesize XXII }&3d $\rightarrow$ 2p\\		
	8 $^{\ast}$& 0.395 & Ar\,{\footnotesize XII }&3d $\rightarrow$ 2p&&   	27 & 0.704&Fe\,{\footnotesize XVIII }&3p $\rightarrow$ 2p&&			  	     & 1.056&Fe\,{\footnotesize XXIII }&3d $\rightarrow$ 2p\\				
	9 $^{\ast}$& 0.399  & S\,{\footnotesize XI }&4d $\rightarrow$ 2p&& 		28 $^{\ast}$& 0.713&O\,{\footnotesize VII }& 5p $\rightarrow$ 1s &&			42 & 1.074&Ne\,{\footnotesize IX }&3p $\rightarrow$ 1s\\ 				
	10 & 0.407 & S\,{\footnotesize XIV }&3p $\rightarrow$ 2s&&			29 & 0.725&Fe\,{\footnotesize XVII } &3s $\rightarrow$ 2p&&				43 & 1.127&Ne\,{\footnotesize IX }&4p $\rightarrow$ 1s\\	
	11 $^{\ast}$& 0.414  & S\,{\footnotesize XII } &4s $\rightarrow$ 2p&& 	     & 0.727&Fe\,{\footnotesize XVII }&3s $\rightarrow$ 2p&&		    		44 & 1.211&Ne\,{\footnotesize X }&3p $\rightarrow$ 1s\\				
	12& 0.420&N\,{\footnotesize VI } &2s $\rightarrow$ 1s (f)&&			30 & 0.739&Fe\,{\footnotesize XVII }&3s $\rightarrow$ 2p&&				45 & 1.277&Ne\,{\footnotesize X }&4p $\rightarrow$ 1s\\					
	13& 0.426&N\,{\footnotesize VI } &2p $\rightarrow$ 1s (i) && 			31 & 0.775&Fe\,{\footnotesize XVIII }&3s $\rightarrow$ 2p&&				46 $^{\ast}$ & 1.314  & Fe\,{\footnotesize XXI }&4d $\rightarrow$ 2p\\	   		     
	14 & 0.431&N\,{\footnotesize VI } &2p $\rightarrow$ 1s (r)&&			    & 0.775&O\,{\footnotesize VIII }&3p $\rightarrow$ 1s&&					47 & 1.352&Mg\,{\footnotesize XI }&2p $\rightarrow$ 1s (r)\\	 		
	15 & 0.436&C\,{\footnotesize VI } &3p $\rightarrow$ 1s&&				32 $^{\ast}$& 0.793  & Fe\,{\footnotesize XVIII }& 3s $\rightarrow$ 2p&&		48  & 1.473&Mg\,{\footnotesize XII }&2p $\rightarrow$ 1s \\					
	16 $^{\ast}$& 0.443 & Ca\,{\footnotesize XII }&3d $\rightarrow$ 2p&&	33 & 0.812&Fe\,{\footnotesize XVII }&3d $\rightarrow$ 2p&&				49 $^{\ast}$& 1.579 & Mg\,{\footnotesize XI }& 3p $\rightarrow$ 1s\\					
	17 $^{\ast}$&0.451 & Ar\,{\footnotesize XIV } &3d $\rightarrow$ 2p &&  	34 & 0.826&Fe\,{\footnotesize XVII }&3d $\rightarrow$ 2p&&				50 $^{\ast}$& 1.659 & Mg\,{\footnotesize XI }& 4p $\rightarrow$ 1s\\
	18 & 0.459&C\,{\footnotesize VI }&4p $\rightarrow$ 1s&&				35 & 0.873&Fe\,{\footnotesize XVIII }&3d $\rightarrow$ 2p&&				51 $^{\ast}$& 1.745  & Mg\,{\footnotesize XII }&3p $\rightarrow$ 1s\\
	19 & 0.500&N\,{\footnotesize VII }&2p $\rightarrow$ 1s&&				36 & 0.905&Ne\,{\footnotesize IX }&2s $\rightarrow$ 1s (f) &&				52 & 1.865&Si\,{\footnotesize XIII }&2p $\rightarrow$ 1s (r)\\ \hline
\enddata
\tablecomments{The asterisk symbols indicate the emission lines that are newly detected in this work.}
\end{deluxetable*}

\section{Response Generation} \label{ssec:response}

\begin{figure}
\gridline{\fig{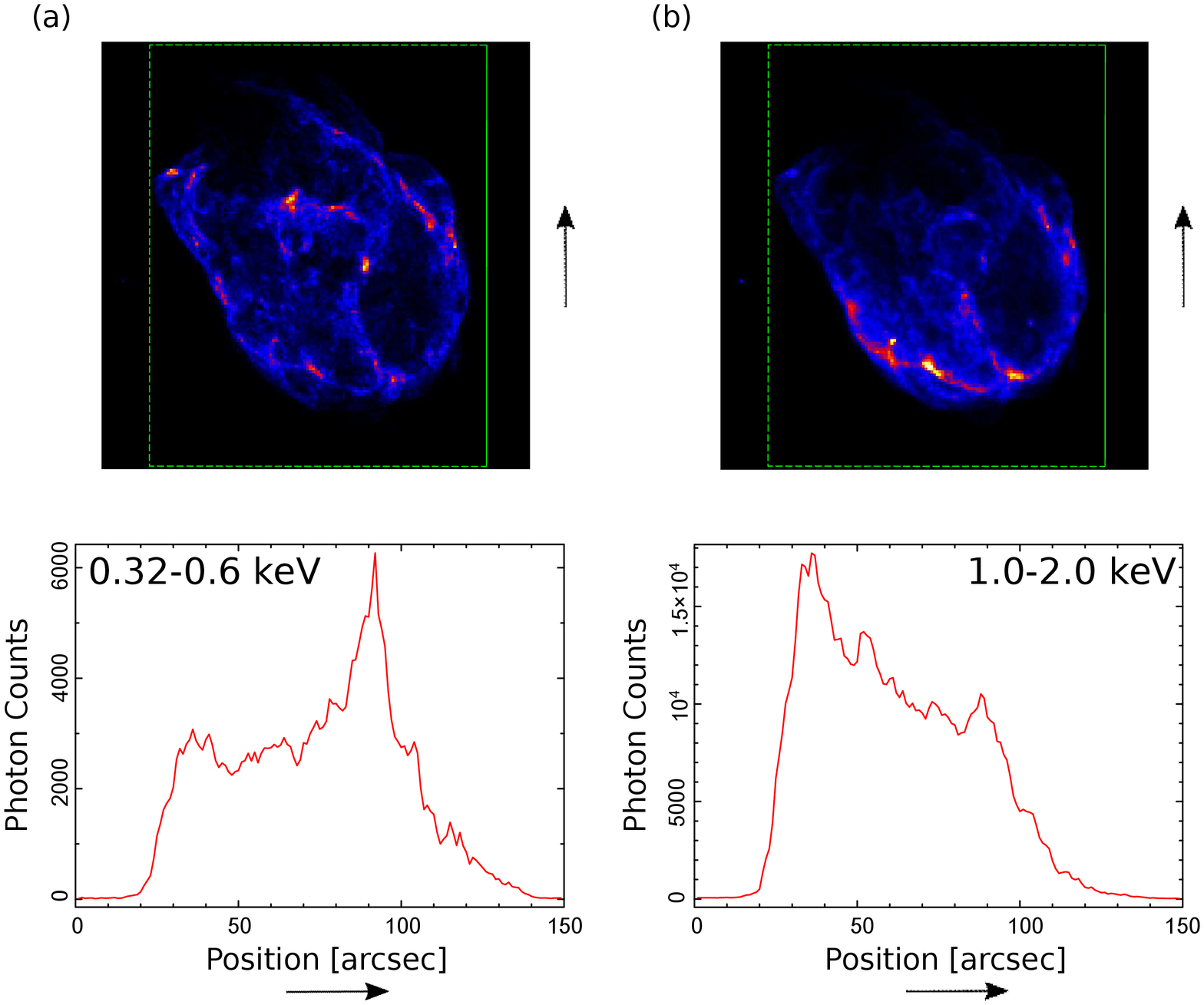}{0.75\textwidth}{}}
\caption{{\it Top}: Chandra ACIS flux images of N132D in the 0.32--0.6\,keV band (a) and 1.0--2.0\,keV band (b). 
The black arrows indicate the dispersion direction of the RGS for Obs.\,ID = 0157161001. 
The green rectangles are where the projection profiles given in the bottom panels are extracted. \ 
{\it Bottom}: Projected profiles of N132D in the energy bands corresponding to the top panel images. 
The profiles are largely different from each other. \label{fig:projection}}
\end{figure}

This section describes how we treat the spectral response, which is crucial 
for the accurate spectral analysis presented in the following sections. 
In general, grating spectrometers are utilized for observing point-like sources, 
and thus a response matrix generated by the standard processes in the {\tt rgsproc} task 
is not applicable to analysis of morphologically asymmetric, extended objects, like N132D. 
Therefore, we convolve the standard response matrix (valid for point-like sources) with 
the projected profiles of the surface brightness along the dispersion axis ((Figure\,\ref{fig:projection}), 
using the publicly-available script 
{\tt ftrgsrmfsmooth}\footnote{https://heasarc.gsfc.nasa.gov/lheasoft/ftools/fhelp/ftrgsrmfsmooth.html}. 
This process is executed for each Obs.ID independently, since the projected profile depends on the roll angle.
We find, furthermore, that the morphology of the SNR varies with energy range, as illustrated in (Figure\,\ref{fig:projection}. 
To account for this issue, we generate projected profiles using archived Chandra images 
(that have the highest available angular resolution) in the five energy ranges of 
0.32--0.6\,keV, 0.6--0.7\,keV, 0.7--0.85\,keV, 0.85--1.0\,keV, and 1.0--2.0\,keV. 
No significant difference is found in the projected profiles of even narrower energy bands. 
Accordingly, we generate convolved response matrices for each of the RGS1 and RGS2 and 
each of the first- and second-order spectra. In the subsequent spectral analysis, 
we apply these responses to the spectra in the corresponding energy bands.

\section{Spectral Modeling} \label{ssec:modeling}

Although we have made the best effort to generate the spectral response for this extended object, 
the roll angle dependence of the line spread function may still cause substantial uncertainties in the modeling. 
We therefore analyze the spectra from each Obs.\,ID independently, rather than performing 
simultaneous spectral fitting, to investigate typical ranges of the systematic uncertainties (and later derive 
the mean values and deviations among the different Obs.\,IDs of the intensities of some strong lines). 
First, we analyze the data from the longest exposure (Obs.\,ID 0157161001) which had a roll angle of 180.55\,deg.
While the second-order spectra have better resolution, 
the first-order spectra cover a wider wavelength range particularly in the soft X-ray band. 
Therefore, we use the second-order spectra for their full bandpass (i.e., 0.66--2.0\,keV) 
and the first-order spectra only when the second order is unavailable (i.e., 0.32--0.68\,keV). 
The XSPEC software version 12.10.0c and the $C$-statistic \citep{Cash1979} on 
unbinned spectra are used in the subsequent spectral analysis.  

Since the previous RGS study indicated the presence of multiple-temperature plasmas \citep{Behar2001}, 
we start spectral modeling with two components of velocity-broadened, collisionally-ionized plasma models, 
{\tt bvrnei}\footnote{https://heasarc.gsfc.nasa.gov/xanadu/xspec/manual/node145.html} 
in the XSPEC package, that can reproduce both CIE and NEI plasmas. The free parameters 
are the electron temperature ($kT_e$), ionization parameter ($n_e t$) 
that is a product of the electron density and time elapsed after shock heating, 
abundances of C, N, O, Ne, Mg, Si, S, Ar, Ca, Fe and Ni relative to the solar values 
of \cite{Wilms2000}, velocity dispersion $\sigma$, and the volume emission measure (VEM). 
The ionization parameters and abundances of each element of the two components are 
tied to each other. 
We also allow the redshift (or blueshift when the value is negative) of each plasma 
component to vary freely, although the resulting values could be dominated by 
the calibration uncertainties in the channel-to-wavelength transformation, 
rather than the real plasma motion. 
The initial plasma temperature ($kT_{\rm init}$) is fixed to 0.01\,keV 
(i.e., assuming that the plasma has been ionizing). 
For the foreground absorption, we introduce the {\tt tbnew} 
models\footnote{https://pulsar.sternwarte.uni-erlangen.de/wilms/research/tbabs/} 
with solar \citep{Wilms2000} and LMC \citep{Russell1992,Schenck2016} abundances 
for the Galactic and LMC components, respectively. The hydrogen column density $N_{\rm H}$ 
of the former is fixed to 6.2 $\times$ 10$^{20}$\,cm$^{-2}$ \citep{Dickey1990}, 
whereas that of the latter is left as a free parameter. 

\begin{figure*}
\gridline{\fig{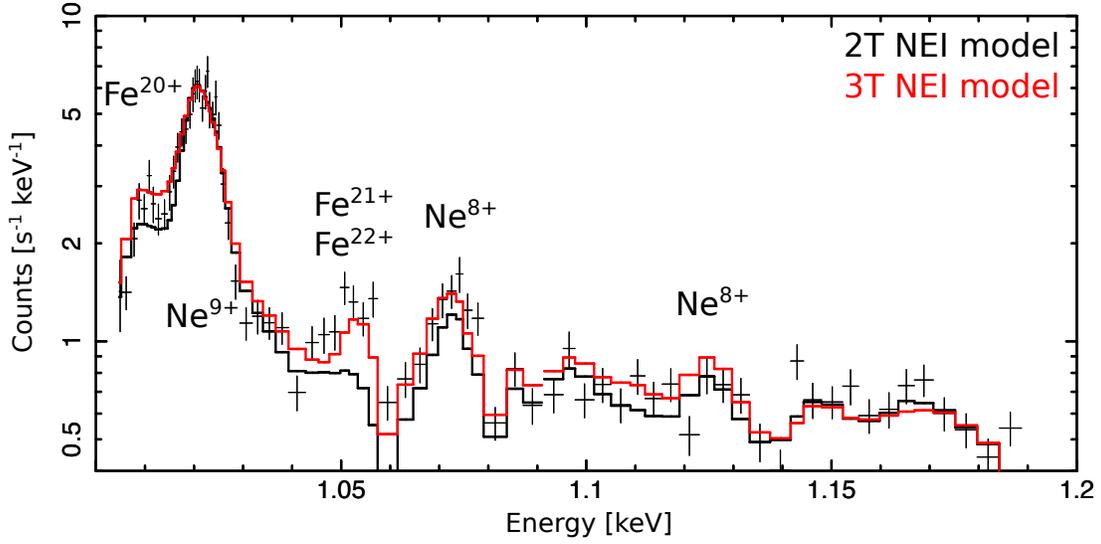}{0.8\textwidth}{}}
\caption{RGS1 second-order spectrum in the 1.0--1.2\,keV band, 
fitted with the two-component (black) and three-component (red) NEI model. 
Only the latter successfully reproduces the emission lines of Fe$^{21+}$ and Fe$^{22+}$ 
around 1.05\,keV. 
\label{fig:2NEI_3NEI}}
\end{figure*}

\begin{figure*}
\gridline{\fig{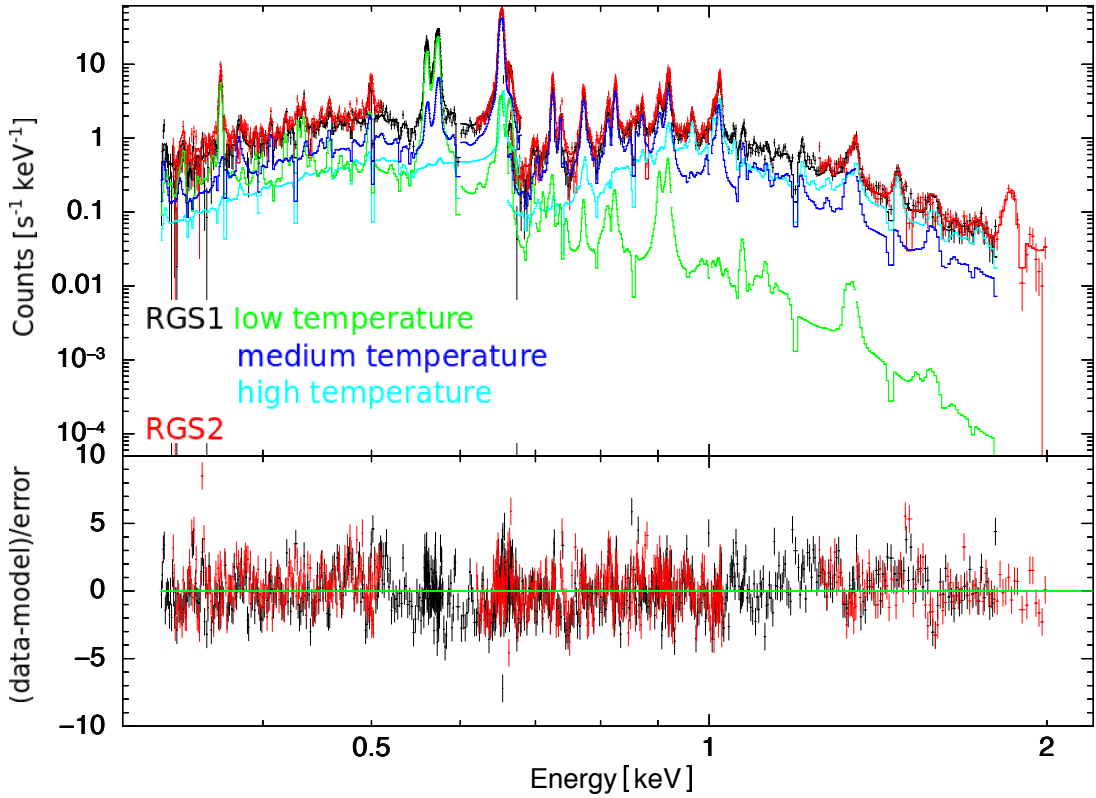}{0.8\textwidth}{}}
\caption{The RGS spectra fitted with the three-component NEI model. 
Black and red are the RGS1 and RGS2, respectively. 
The spectra below 0.68\,keV are from the first-order data, 
whereas those above 0.66\,keV are from the second-order data.
The colored solid curves indicate the contribution of the individual components,  
low-$T_e$ (green), medium-$T_e$ (blue), and high-$T_e$ (cyan) plasmas, to the RGS1 data.
The bottom panel shows the residual from the best-fit model.
\label{fig:3NEI}}
\end{figure*}

With these models and assumptions, we obtain the best-fit result with 
the electron temperatures of $\sim$\,0.31\,keV and $\sim$\,0.91\,keV, ionization parameter of 
$\sim$\,1.3 $\times$ $10^{11}$\,cm$^{-3}$\,s, and $C$-stat/d.o.f.\ of 10962/7567.
We find a significant discrepancy between the model and data particularly around 1.05\,keV;  
the model (the black solid curve in Figure\,\ref{fig:2NEI_3NEI}) fails to reproduce the emission lines 
of Fe$^{21+}$ and Fe$^{22+}$ at this energy. 
This indicates the presence of even hotter plasma where the Fe is more highly ionized. 
We thus add another NEI component with a different electron temperature, 
but with the ionization parameter and element abundances tied to those of the other components.
We obtain a successful fit to these emission lines as well as the broadband spectra 
(Figure\,\ref{fig:2NEI_3NEI} red solid curve and Figure\,\ref{fig:3NEI}, respectively) 
with an improved $C$-stat/d.o.f.\ of 10426/7563. 
The best-fit parameters are given in Table\,\ref{tab:3NEI_para}. 
The constrained ionization parameter is significantly lower than 10$^{12}$\,cm$^{-3}$\,s, indicating that the plasma is in the NEI. 
We also attempt to model the spectra with three-component CIE plasmas ({\tt bvapec} model in XSPEC) 
and reach the results given in Table \ref{tab:NEI_CIE}, confirming that the NEI (ionizing) plasma model 
indeed provides a better fit. 
No improvement is obtained when the initial plasma temperature $kT_{\rm init}$ of {\tt bvrnei} is 
fixed to 3.0\,keV assuming a recombining state.

\begin{deluxetable*}{llll|}
\tablecaption{The best-fit parameters of the three-temperature NEI model. \label{tab:3NEI_para}}
\tablewidth{0pt}
\tablehead{
\colhead{Parameters}&					& \colhead{} & \colhead{}}
\startdata
$N\rm_H$ 					&(10$^{20}$ cm$^{-2}$)	&6.8$^{+0.1}_{-0.4}$\\
$kT_{e,{\rm low}}$ 				&(keV) 				&$0.200^{+0.004}_{-0.005}$\\ 
VEM$_{\rm low}$    		 		&(10$^{60}$ cm$^{-3}$) 	&$1.31^{+0.03}_{-0.04}$ \\
$\sigma_{\rm low}$				&(km s$^{-1}$) 			&$438 \pm 34$\\
$v_{\rm low}$ 					&(km s$^{-1}$) 			&$559 \pm 18$\\
$kT_{e,{\rm med}}$ 				&(keV) 				&$0.563^{+0.01}_{-0.005}$\\
VEM$_{\rm med}$	        		        & (10$^{60}$ cm$^{-3}$) 	&$1.52^{+0.02}_{-0.03}$ \\
$\sigma_{\rm med}$				&(km s$^{-1}$) 			&$445^{+21}_{-20}$\\
$v_{\rm med}$  				&(km s$^{-1}$) 			&$183 \pm 11$\\
$kT_{e,{\rm high}}$ 			 	&(keV) 				&$1.36^{+0.04}_{-0.02}$\\
VEM$_{\rm high}$				& (10$^{60}$ cm$^{-3}$) 	&$0.93^{+0.05}_{-0.03}$ \\
$\sigma_{\rm high}$				&(km s$^{-1}$) 			&0 (fixed)\\
$v_{\rm high}$  					&(km s$^{-1}$) 			&$-639 \pm 27$\\
C      						&					&0.26$^{+0.02}_{-0.01}$\\
N    							& 					&0.172$^{+0.009}_{-0.010}$\\
O     							&					&0.34$^{+0.01}_{-0.02}$\\
Ne     						& 					&0.51$^{+0.02}_{-0.01}$\\
Mg     						& 					&0.49$^{+0.03}_{-0.02}$\\
Si     							& 					&$0.59 \pm 0.05$\\
S      							&					&$0.57^{+0.06}_{-0.03}$\\
Ar      						&					&$0.75^{+0.09}_{-0.07}$\\
Ca      						&					&$0.04^{+0.12}_{-0.04}$\\
Fe      						&					&$0.411^{+0.014}_{-0.007}$\\
Ni      						&					&$0.71^{+0.11}_{-0.09}$\\
$n_et$						&($10^{10}$\,cm$^{-3}$\,s ) &$9.8^{+0.3}_{-0.5}$\\\hline
$\it C$-statistics/d.o.f. && 10426/7563\\ \hline
\enddata
\tablecomments{The volume emission measure (VEM) is defined as $\int n_e n_H dV$, 
where $n_e$ is the electron density, $n_H$ the hydrogen density, and $V$ the volume of the source.
$v$ is the line-of-sight bulk velocity of plasmas inferred from the Doppler shift of 
the emission lines. The velocity dispersion of the high temperature component 
($\sigma_{\rm high}$) is fixed because it is found to be consistent with 0.}
\end{deluxetable*}

\begin{deluxetable*}{lll}
\tablecaption{Comparison of the best-fit parameters of the NEI and CIE cases. \label{tab:NEI_CIE}}
\tablewidth{0pt}
\tablehead{
\colhead{} & \colhead{Ionizing} & \colhead{CIE}}
\startdata
$ kT_{e,{\rm low}}$ (keV) &  	$0.200^{+0.004}_{-0.005}$&  	$0.150 \pm 0.003$  \\
$ kT_{e,{\rm med}}$ (keV) &  	$0.563^{+0.01}_{-0.005}$& 	 $0.379\pm 0.004$   \\
$ kT_{e,{\rm high}}$ (keV) &  	$1.36^{+0.04}_{-0.02}$   &  	$0.896\pm 0.007$  \\
$ kT_{\rm init}$ (keV) &  		0.01 (fixed) & ---  \\
$n_et$ (\,cm$^{-3}$\,s) & 		$(9.8^{+0.3}_{-0.5})\times 10^{10}$ & ---   \\  \hline
$C$-stat/d.o.f. & 10426/7563   &  10570/7564  \\  \hline
\enddata
\end{deluxetable*}

\begin{figure}
\gridline{\fig{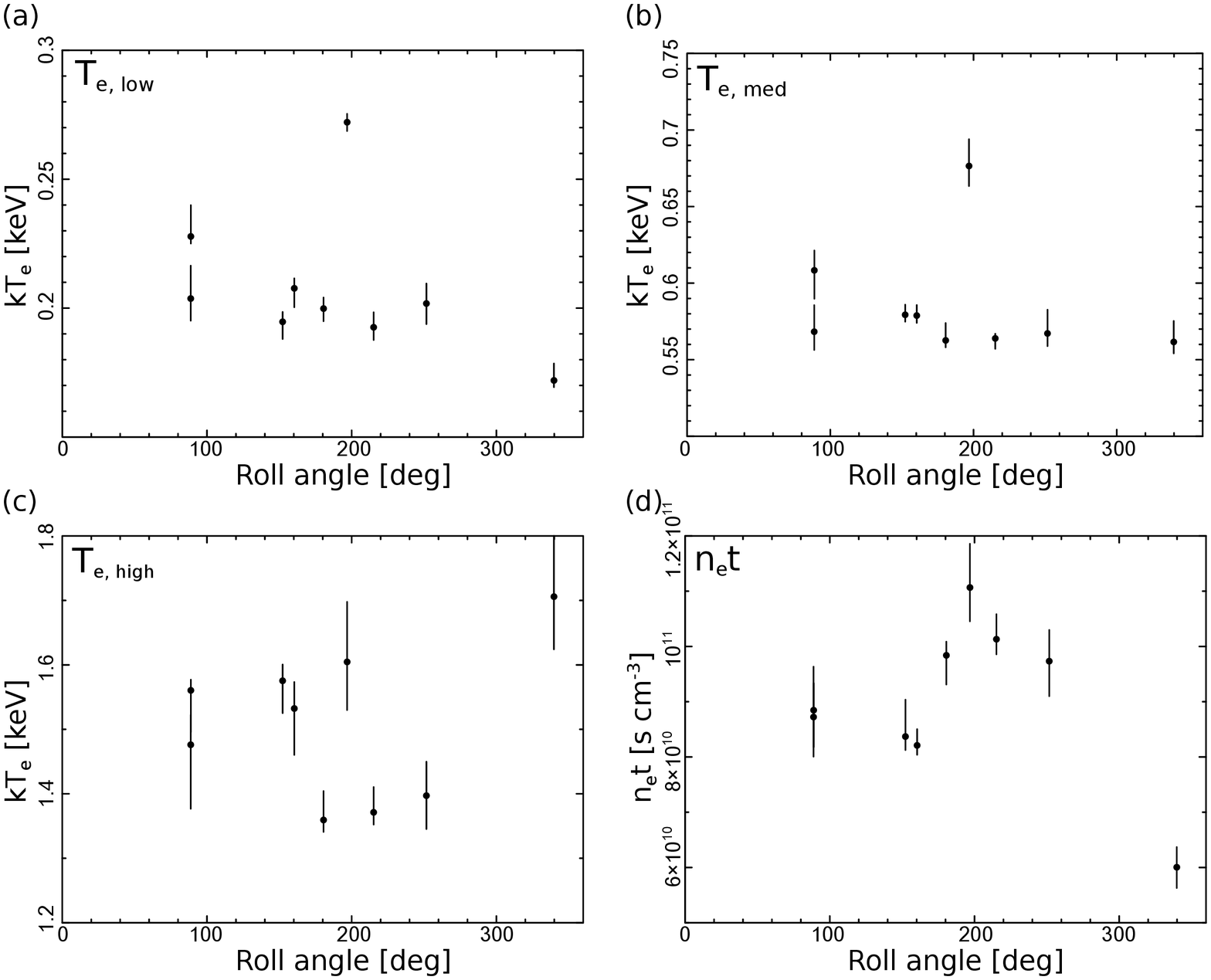}{0.8\textwidth}{}}
\caption{The best-fit electron temperatures and ionization timescale obtained from different observations.  
\label{fig:roll}}
\end{figure}

\begin{figure*}
\gridline{\fig{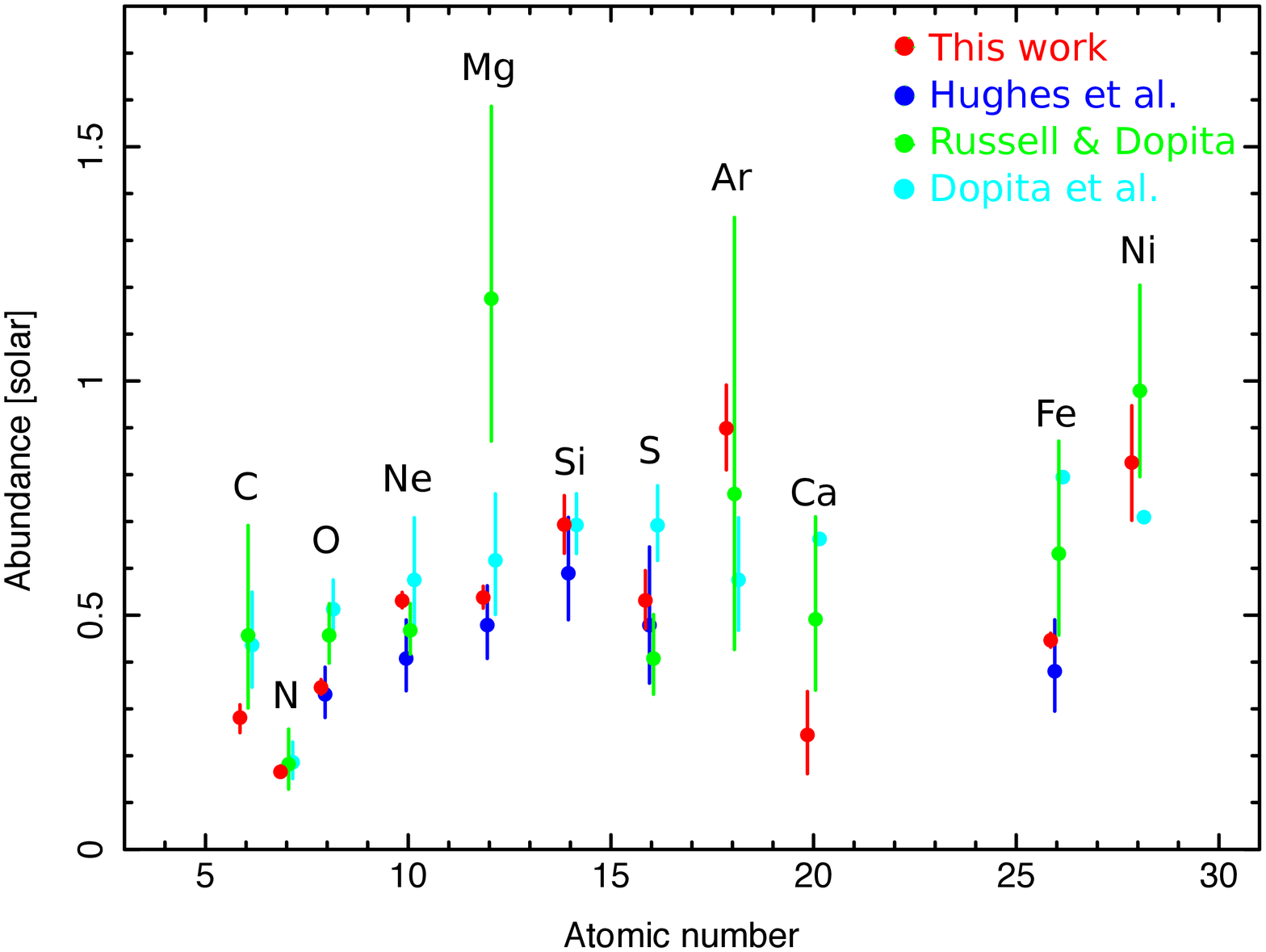}{0.6\textwidth}{}}
\caption{Elemental abundances (relative to the solar values of \cite{Wilms2000})
as a function of atomic number derived from our spectral analysis (red). 
The previous measurements by \cite{Hughes1998}(blue), \cite{Russell1992} (green), and \cite{Dopita2019}(cyan)
are also normalized by the solar abundances of \cite{Wilms2000} and 
plotted for comparison. 
\label{fig:abund}}
\end{figure*}

Next we analyze the data from the other on-axis observations to investigate how the results depend on 
the Obs.\,ID with different spacecraft roll angles. Applying the same data reduction and spectral models, 
we obtain the electron temperatures and ionization timescale given in Figure\,\ref{fig:roll} for each observation. 
Except for a few cases, the values are consistent within $\sim$\,10\% among the observations. 
We find that the Obs.\,ID 0157350201 data 
(roll angle = 196.82) give relatively high electron temperatures 
$kT_{e,{\rm low}}$ $\sim$ 0.27\,keV, $kT_{e,{\rm med}}$ $\sim$ 0.67\,keV. 
However, even if we fit the data with the fixed values of $kT_{e,{\rm low}}$ = 0.2\,keV and 
$kT_{e,{\rm med}}$ = 0.6\,keV (that are comparable to the values for the other Obs.\,IDs), 
the $C$-stat value only increases from 9582 to 9645 and the model still fits the data quite well. 
We therefore conclude that, regardless of the roll angle, the three-temperature model with 
$kT_e$ of $\sim$\,0.2\,keV, $\sim$\,0.6\,keV, and $\sim$\,1.5\,keV gives a reasonable 
approximation of the observed spectrum that is extracted from the entire SNR. 
We also find that the elemental abundances have small variation among the Obs.\,IDs 
and their mean values (Figure\,\ref{fig:abund}) are comparable to the previous measurements 
of the LMC ISM \citep{Hughes1998,Russell1992,Dopita2019}. 
This suggests the swept-up ISM origin 
of the soft X-rays from this SNR, which is also supported by its shell-like morphology 
at the energies below 2.0 keV (Figure\,\ref{fig:projection}). 
Strictly speaking, the Ca abundance is slightly lower than the typical value for the LMC.
This may imply that a fraction of this element is depleted by dust grains, 
as is the case in e.g., Cas~A \citep{DeLooze2017}. An alternative possibility is that 
the L-shell line emissivity for Ca is inaccurate in the current atomic database.

%%図修正 20200707 -> 2019年の論文に 'updated' solar abundance は少し違和感があったので微修正 

Assuming a spherically symmetric shell with an outer radius of $R \sim 1'$ and a thickness of $R/12$ 
for the swept-up ISM, we obtain the plasma volume to be $8.7 \times 10^{58}$\,cm$^3$ 
at the distance to the LMC. 
The VEM of each component derived from our spectral modeling (Table\,\ref{tab:3NEI_para}), therefore, 
corresponds to the post-shock proton density ($n_{\rm H}$) and swept-up ISM mass ($M_{\rm ISM}$) of 
3.6\,cm$^{-3}$ and 260\,$M_{\odot}$ for the low-$T_{\rm e}$ component, 
3.9\,cm$^{-3}$ and 280\,$M_{\odot}$ for the medium-$T_{\rm e}$ component, and
3.0\,cm$^{-3}$ and 220\,$M_{\odot}$ for the high-$T_{\rm e}$ component. 
These values are almost consistent with the previous measurements \citep[e.g.,][]{Hughes1998,Williams2006}, 
although our estimate is based on some over-simplified assumptions 
(e.g., spherically symmetric shell, constant density over the whole SNR).

\begin{figure}
\gridline{\fig{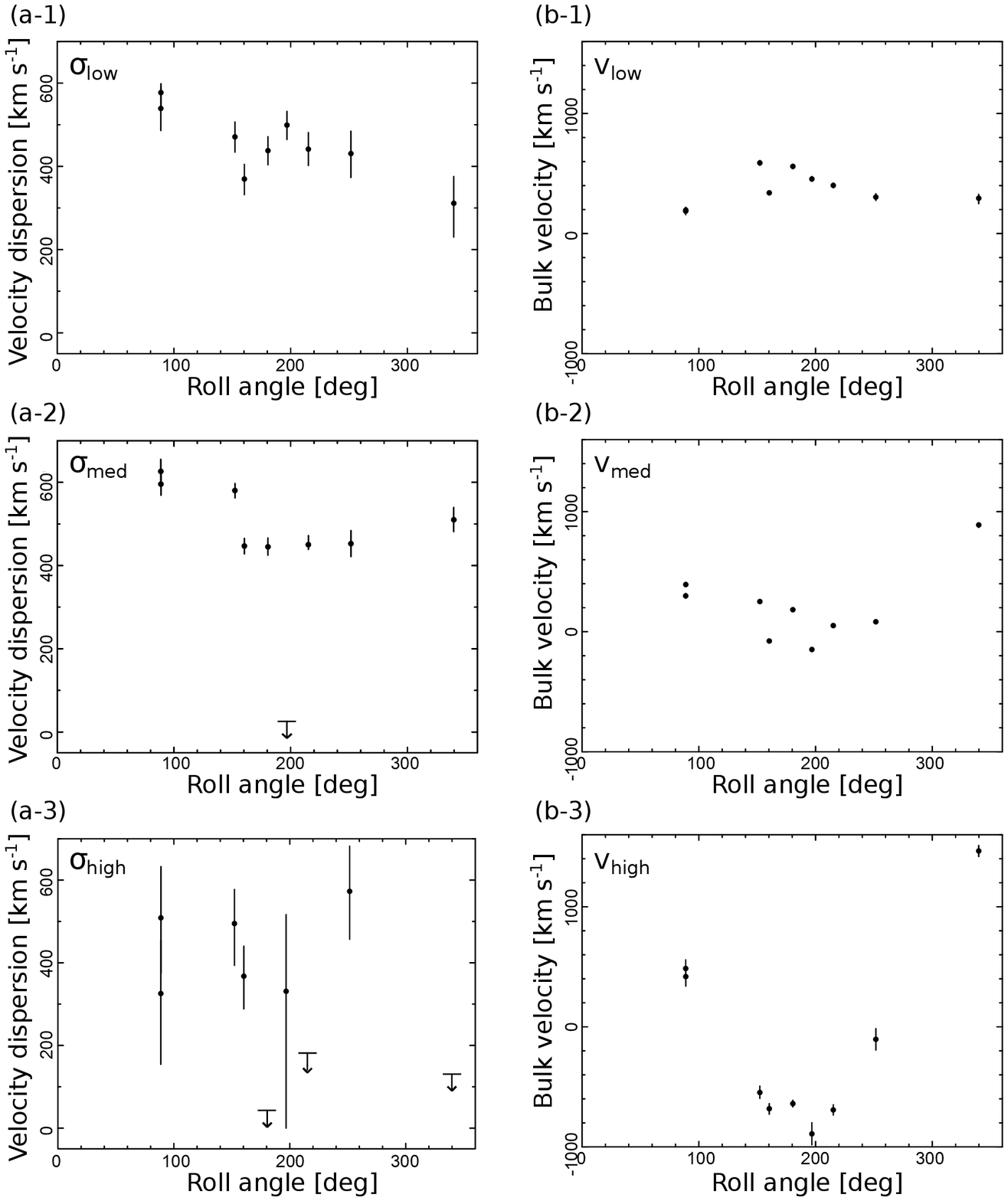}{0.8\textwidth}{}} 
\caption{The line-of-sight velocity dispersion and bulk velocity for different observations. 
The values are subject to the uncertainties in the calibration of the relationship 
between the detector position and wavelength. 
\label{fig:roll_v_z}}
\end{figure}

Figure\,\ref{fig:roll_v_z} shows the line-of-sight velocity dispersion and bulk velocity
for each Obs.\,ID, converted from the spectral line broadening and shift, respectively.
The typical velocity dispersion is 400--500\,km\,s$^{-1}$, consistent with the Hitomi
measurement using the S\,{\footnotesize XV} emission \citep{2018PASJ7016H}.
The redshift of the low- and medium-temperature components is comparable to
the radial velocity of the LMC ISM at the region around this SNR
\citep[$275\pm 4$ km\,s$^{-1}$][]{Vogt2011}.
However, the dispersion among the observations is much larger than the typical
statistical errors, implying that our measurements are dominated by
the calibration uncertainties in the channel-to-wavelength transformation
and/or incompleteness of our response generation for an extended source,
rather than the real plasma motion. Although an accurate measurement
of the line broadening and shift is crucial to understand the SNR kinematics,
this is out of scope of the present work.

\section{Line Diagnostics and Discussion} \label{sec:diagnostics}

In the previous subsection, our analysis has indicated that the plasma is in the NEI (ionizing) 
state ($n_et\approx 10^{11}$\,cm$^{-3}$\,s),
while \cite{Behar2001} did not distinguish between a CIE and NEI plasma.
Here we reconfirm this result, based on more detailed spectral diagnostics 
that utilizes the well-resolved emission lines of O and Ne. 
The intensities of the resonance, forbidden, and $n$ = 3 $\rightarrow$ 1 transitions 
(hereafter ``$r$'', ``$f$'', and ``He$\beta$'', respectively) of O\,{\footnotesize VII} and 
Ne\,{\footnotesize IX}, and the O\,{\footnotesize VIII} and Ne\,{\footnotesize X} 
Ly$\alpha$ lines are directly measured with the following procedure\footnote{A similar procedure 
was applied to the Fe K-shell emission from the Perseus Cluster by \cite{hitomicollaboration2018} 
to estimate the effect of the resonance scattering in the cool core region of this object.}. 
First we modify one of the model data files that the XSPEC package includes, named 
{\tt apec\_v3.0.9\_nei\_line.fits}\footnote{Usually found in the directory of heasoft-6.24/spectral/modelData.}, 
such that all the emission lines mentioned above are deleted (hereafter ``modified NEI''). 
For each Obs.\,ID, the best-fit three-temperature NEI model obtained in Section\,\ref{ssec:modeling} is 
replaced with a model consisting of the modified NEI plus eight Gaussians for 
the $r$, $f$, and He$\beta$ emission of O\,{\footnotesize VII} and Ne\,{\footnotesize IX}, 
and the Ly$\alpha$ emission of O\,{\footnotesize VIII} and Ne\,{\footnotesize X}. 
All the parameters (e.g., electron temperatures, elemental abundances) of the modified NEI 
components as well as the foreground absorption column are fixed to the best-fit values of 
the original three-temperature model. Then we fit the Gaussian components, 
obtaining the line fluxes for each Obs.\,ID in Table\,\ref{tab:line_intensity}. 
Finally, we calculate the exposure-weighted mean flux ($F$) 
and their statistical errors ($\sigma_{F{\rm , stat}}$), as well as 
the standard deviations among the different Obs.\,IDs ($\sigma_{F{\rm , SD}}$) as follows: 
\begin{equation}
\begin{split}
F &= \frac{\sum_{i} F_i \times t_i}{\sum_{i} t_i} \\
\sigma_{F{\rm , stat}} &= \sqrt {\frac{(\sigma_i)^2 \times t_i}{\sum_i t_i}}    \\
\sigma_{F{\rm , SD}} &= \sqrt{\frac{\sum_{i} (F_i - F_{\rm mean})^2}{9}},  \\
\end{split}
\end{equation}
where $F_i$, $\sigma_i$, and $t_i$ are the line flux, its statistical error, 
and exposure of the observation $i$. 
The resulting values are given in Table\,\ref{tab:line_intensity}.

\begin{deluxetable*}{llllllllll}
\tablecaption{The intensities of the O and Ne emission. \label{tab:line_intensity}}
\tablewidth{0pt}
\tablehead{
\colhead{Obs. ID}&\colhead{exposure time} & \colhead{O r}& \colhead{O f} & \colhead{O He$\beta$} & \colhead{O Ly$\alpha$}& \colhead{Ne r}& \colhead{Ne f} & \colhead{Ne He$\beta$} & \colhead{Ne Ly$\alpha$}\\
\colhead{}&\colhead{ksec }&\colhead{$10^{-3}$}    &\colhead{$10^{-3}$}&\colhead{$10^{-3}$}&\colhead{$10^{-2}$}&\colhead{$10^{-3}$}    &\colhead{$10^{-3}$}&\colhead{$10^{-4}$}&\colhead{$10^{-3}$}}
\startdata
0125100201&	13.63&	6.99 $\pm$ 0.13&		4.58 $\pm$ 0.18&		1.31 $\pm$ 0.06&	1.31 $\pm$ 0.01&	2.73 $\pm$ 0.09&	1.34 $\pm$ 0.08&	5.46 $\pm$ 0.79&		2.34 $\pm$ 0.08\\
0157160601&	26.20&	7.94 $\pm$ 0.19&		5.30 $\pm$ 0.10&		1.26 $\pm$ 0.03&	1.36 $\pm$ 0.01&	2.64 $\pm$ 0.07&	1.40 $\pm$ 0.05&	4.57 $\pm$ 0.54&		2.37 $\pm$ 0.06\\
0157160801&	29.09&	9.17 $\pm$ 0.12&		5.93 $\pm$ 0.13&		1.29 $\pm$ 0.03&	1.52 $\pm$ 0.01&	2.79 $\pm$ 0.08&	1.60 $\pm$ 0.05&	4.55 $\pm$ 0.49&		2.47 $\pm$ 0.06\\
0157161001&	29.91&	8.09 $\pm$ 0.10&		5.61 $\pm$ 0.13&		1.22 $\pm$ 0.04&	1.40 $\pm$ 0.01&	2.79 $\pm$ 0.06&	1.33 $\pm$ 0.07&	4.97 $\pm$ 0.42&		2.39 $\pm$ 0.06\\ 
0157360201&	15.02&	8.48 $\pm$ 0.17&		5.69 $\pm$ 0.17&		1.38 $\pm$ 0.06&	1.40 $\pm$ 0.02&	2.43 $\pm$ 0.10&	1.46 $\pm$ 0.08&	5.98 $\pm$ 1.23&		2.52 $\pm$ 0.08\\
0157360301&	28.66&	8.49 $\pm$ 0.12&		5.79 $\pm$ 0.09&		1.25 $\pm$ 0.03&	1.49 $\pm$ 0.01&	2.67 $\pm$ 0.07&	1.59 $\pm$ 0.07&	6.49 $\pm$ 0.56&		2.44 $\pm$ 0.06\\
0157360501&	14.92&	7.99 $\pm$ 0.27&		5.72 $\pm$ 0.16&		1.31 $\pm$ 0.07&	1.56 $\pm$ 0.01&	2.78 $\pm$ 0.10&	1.59 $\pm$ 0.10&	4.10 $\pm$ 0.78&		2.53 $\pm$ 0.08\\
0129341301&	20.34&	8.10 $\pm$ 0.15&		5.57 $\pm$ 0.15&		1.43 $\pm$ 0.04&	1.41 $\pm$ 0.01&	2.64 $\pm$ 0.07&	1.48 $\pm$ 0.07&	4.46 $\pm$ 0.61&		2.43 $\pm$ 0.05\\
0137551101&	15.04&	8.55 $\pm$ 0.15&		5.90 $\pm$ 0.13&		1.37 $\pm$ 0.05&	1.44 $\pm$ 0.02&	2.63 $\pm$ 0.12&	1.57 $\pm$ 0.07&	5.44 $\pm$ 0.81&		2.43 $\pm$ 0.07\\
Mean           &           &	8.27&	  				5.60&					1.30&			1.44&			2.69&			1.48&  	         	5.11&				2.43\\
Statistical error&	        &\multicolumn{1}{r}{0.17}&\multicolumn{1}{r}{0.16}&\multicolumn{1}{r}{0.05}&\multicolumn{1}{r}{0.01}&\multicolumn{1}{r}{0.08}&\multicolumn{1}{r}{0.07}&\multicolumn{1}{r}{0.67}&	\multicolumn{1}{r}{0.07}\\
Standard deviation&  &\multicolumn{1}{r}{0.56}&	\multicolumn{1}{r}{0.39}&\multicolumn{1}{r}{0.07}	&\multicolumn{1}{r}{0.08}&\multicolumn{1}{r}{0.11}&\multicolumn{1}{r}{0.10}&\multicolumn{1}{r}{0.74}&\multicolumn{1}{r}{0.06}\\ \hline
\enddata
\tablecomments{In units of photons cm$^{-2}$s$^{-1}$}
\end{deluxetable*}

\begin{deluxetable*}{lllllllllll|l|l|l|}
\tablecaption{The mean line intensity ratios. \label{tab:intensity_ratio}}
\tablewidth{0pt}
\tablehead{
\colhead{}&\colhead{He$\beta$ / He$\alpha$} & \colhead{Ly$\alpha$ / He$\alpha$}& \colhead{$f$/$r$}}
\startdata
O &0.094 $\pm$ 0.004&	1.04 $\pm$ 0.02&	 0.68 $\pm$ 0.02	\\
Ne &0.12 $\pm$ 0.02&	0.58 $\pm$ 0.02&	0.55 $\pm$ 0.03 \\
\enddata
\end{deluxetable*}

\begin{figure*}
\gridline{\fig{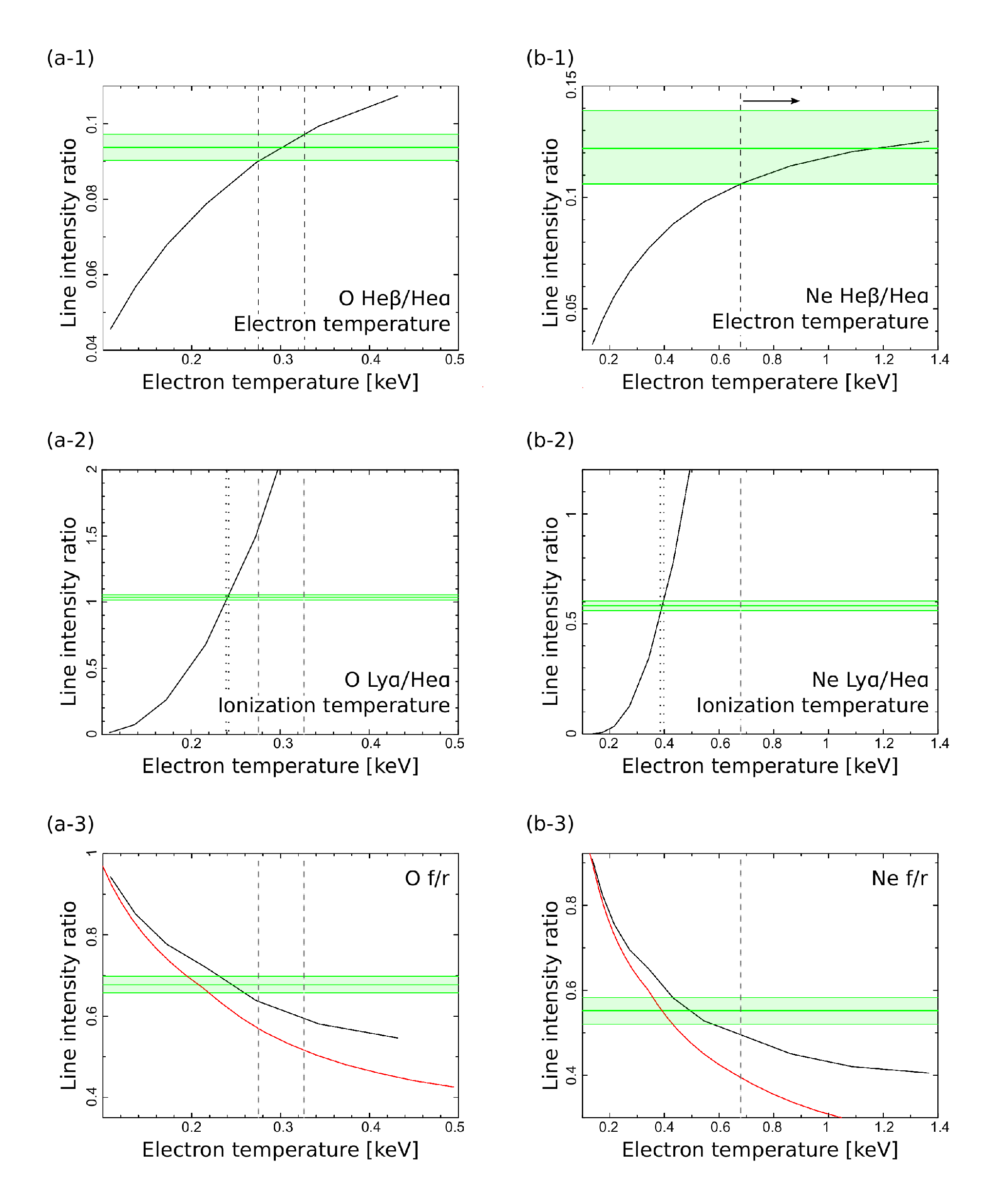}{0.9\textwidth}{}}
\caption{Comparison between the theoretical and observational ratios of 
the diagnostic emission lines. 
(a-1): O He$\beta$/He$\alpha$ ratio that represents the electron temperature, 
where He$\alpha$ is defined as the sum of O\,{\footnotesize VII} resonance and forbidden lines. 
(a-2): O Ly$\alpha$/He$\alpha$ ratio that represents the ionization temperature. 
(a-3): O forbidden-to-resonance line ratio.
Panels (b-1), (b-2), and (b-3) are the same as (a-1), (a-2), and (a-3), but for Ne. 
The black curves are the theoretically-predicted ratios as a function of the electron temperature, 
derived using PyAtomDB assuming CIE. 
The red curves in panels (a-3) and (b-3) are those expected for NEI plasmas with 
$n_et = 10^{11}$\,cm$^{-3}$\,s. 
The green areas indicate the observed line ratios. 
The dashed and dotted lines indicate the constrained electron and ionization temperatures, respectively. 
\label{fig:ratio}}
\end{figure*}

The line ratios given in Table\,\ref{tab:intensity_ratio} offer useful diagnostics for the plasma condition. 
The He$\beta$/He$\alpha$ ratio (where He$\alpha$ is defined to be the sum of the resonance and forbidden lines) 
almost solely depends on the electron temperature, whereas the Ly$\alpha$/He$\alpha$ ratio reflects 
the charge balance between the H-like and He-like ions (so-called ``ionization temperature''). 
The forbidden-to-resonance ratio depends not only on both electron and ionization temperatures 
but also on external effects, such as resonance scattering \citep[e.g.,][]{Miyata2008} 
and charge exchange \citep[e.g.,][]{Uchida2019}. 
The black curves in Figure\,\ref{fig:ratio} (a-1) and (b-1) are the theoretically predicted 
He$\beta$/He$\alpha$ ratios of O and Ne (for CIE plasmas), 
as a function of the electron temperature, 
derived using {\tt PyAtomDB}\footnote{https://github.com/AtomDB/pyatomdb}. 
The observed line ratios (green areas) are compared with these theoretical values, 
constraining the characteristic electron temperatures to 0.28--0.33\,keV for O and $>$\,0.68\,keV for Ne. 
We also estimate the electron temperatures based on the He$\beta$/$r$ or He$\beta$/$f$ ratio, 
instead of He$\beta$/He$\alpha$, and find that the constrained values do not change significantly. 
Similarly, the observed Ly$\alpha$/He$\alpha$ ratios (that represent the ionization temperature) are 
compared with the theoretical ratios for CIE plasmas in Figure\,\ref{fig:ratio} (a-2; for O) and (b-2; for Ne). 
The obtained ionization temperature ($\sim$\,0.24\,keV for O; $\sim$\,0.39\,keV for Ne) is lower than the 
characteristic electron temperature of each element, confirming that the plasma is in the ionizing state.

\begin{figure}
\gridline{\fig{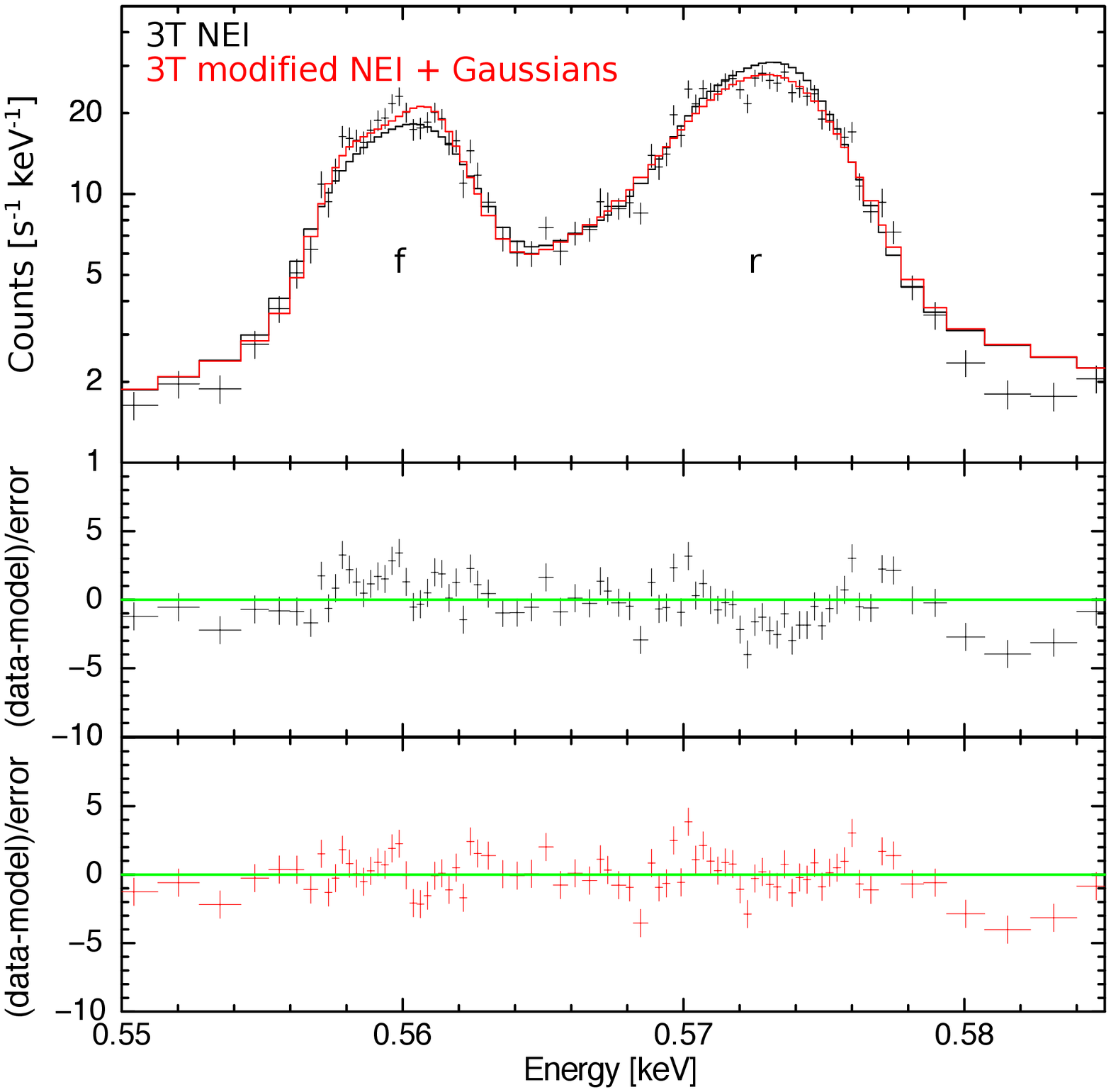}{0.6\textwidth}{}}
\caption{
The RGS1 spectrum of the O\,{\footnotesize VII} He$\alpha$ emission. 
The black curve in the top panel is the best-fit three-temperature NEI model, 
where not only the electron temperature but also the ionization timescale 
and O abundance of each component are independently fitted. 
The red curve is the model consisting of modified NEI and Gaussians (see text). 
The middle and bottom panels show the residual between the model and data 
corresponding to the black and red curves in the top panel, respectively. 
\label{fig:f_r_hikaku}}
\end{figure}

The black curve in Figure\,\ref{fig:ratio} (a-3) shows the O\,{\footnotesize VII} $f$/$r$ ratio 
as a function of the electron temperature that is expected for a CIE plasma. 
We find that the observed $f$/$r$ ratio (green area) corresponds to 0.23--0.26\,keV, significantly lower 
than the characteristic electron temperatures determined by the He$\beta$/He$\alpha$ ratio (panel a-1). 
The gap between the two temperatures becomes even larger, when the observed $f$/$r$ ratio 
is compared with the theoretical values for an ionizing plasma with $n_et = 10^{11}$\,cm$^{-3}$\,s 
(red curve; the best-fit ionization parameter for the three-component NEI model). 
Similar results are obtained for the Ne\,{\footnotesize IX} $f$/$r$ ratio (Figure\,\ref{fig:ratio} b-3). 
These discrepancies could be just due to the complexity of the multiple-temperature plasmas,
whose spectrum cannot be perfectly reproduced by our approximated three-temperature model. 
We find, however, that the discrepancies remain even if the ionization timescale and 
elemental abundances (in addition to the electron temperature) are untied among 
the three components (Figure\,\ref{fig:f_r_hikaku}).
It would, therefore, be worth discussing another possibility that the observed $f$/$r$ ratios 
are modified by other effects, such as resonance scattering and charge exchange.

\subsection{Resonance Scattering}

The effect of resonance scattering is usually negligible in SNRs, since their plasmas are optically thin. 
However, if the column density of certain ions is relatively high, some of the resonance line photons 
can be scattered out from the line of sight when propagating in the X-ray emitting plasma itself 
\citep[e.g.,][]{Miyata2008,Amano2020}. 
Since N132D exhibits a dense SNR shell in the soft X-ray band \citep{Behar2001,Borkowski2007} 
and its geometry is known to depart from spherical symmetry \citep[e.g.,][]{Law2020},
it is indeed possible that resonance scattering has taken place and modified the observed $f$/$r$ ratio. 
Based on this assumption, we revisit the full-band RGS spectra and conduct the following analysis. 
In addition to the best-fit three-temperature model obtained in Section 4, we include two Gaussians 
at 0.574\,keV and 0.922\,keV (corresponding to the O\,{\footnotesize VII} and Ne\,{\footnotesize IX} 
resonance lines, respectively) in the spectral fitting by allowing their normalization to have negative values. 
This analysis yields the $C$-stat value of 10412 (reduced from 10426) and 
Gaussian normalizations (the mean of the nine Obs.\,IDs) of 
--4.7 ($\pm$ 1.5) $\times 10^{-4}$ photons\,cm$^{-2}$\,s$^{-1}$ for O\,{\footnotesize VII} $r$ and 
--1.6 ($\pm$ 0.8) $\times 10^{-4}$ photons\,cm$^{-2}$\,s$^{-1}$ for Ne\,{\footnotesize IX} $r$. 
Both are indeed smaller than zero, as implied by Figure\,\ref{fig:ratio} a-3 and b-3. 
If the optical depth ($\tau$) for the line photons is significantly smaller than unity, 
the relationship between the scattered and observed photon flux 
($\Delta I$ and $I_{\rm obs}$, respectively) is given as $\Delta I \approx \tau I_{\rm obs}$. 
Thus, attributing the negative Gaussian normalizations to the resonance scattering, 
we obtain $\tau _{{\rm O,}r} = 0.057 \pm 0.018$ and $\tau _{{\rm Ne,}r} = 0.059 \pm 0.030$.

Theoretically, the cross section of the resonance scattering is given as 
$\sigma \sim 1.3 \times 10^{-9} \cdot f / E \cdot (m_i / kT_i)^{1/2}$\,cm$^2$, 
where $f$, $E$, $m_i$, and $T_i$ are the oscillator strength of the line transition, photon energy in keV, 
and the mass and temperature of the ions, respectively \citep[e.g.,][]{Miyata2008}. 
The optical depth for the resonance scattering can, therefore, be expressed as 
\begin{equation}
\begin{split}
\tau &= \sigma n_z L \\
&= 1.3 \times 10^{-9} \cdot \frac{f}{E} \left(\frac{m_i}{kT_i}\right)^{1/2} A_Z \left(\frac{n_z}{n_{Z}}\right) \left(\frac{n_Z}{n_{\rm H}}\right)_\odot n_{\rm H} L, 
\end{split}
\end{equation}
where $n_z$ is the ion density (that is assumed to be constant along the line of sight), 
$L$ is the physical depth of the plasma, 
$A_Z$ is the abundance of the element $Z$ relative to the solar value of \cite{Wilms2000}, 
$n_Z$ is the element density, and $n_{\rm H}$ the hydrogen density; 
thus $n_z/n_Z$ and $(n_Z/n_{\rm H})_{\odot}$ represent the ion fraction of the ion $z$ 
and the solar abundance of the element $Z$, respectively.
Table\,\ref{tab:tau_rs} gives the coefficients we are now interested in. 
We refer to the AtomDB database (available via PyAtomDB) to obtain the oscillator strengths and ion fractions. 
For the latter, the ionization temperatures constrained in Figure\,\ref{fig:ratio} (a-2) and (b-2) are applied.
Since the resonance line of O\,{\footnotesize VII} is dominated by the low-$T_{\rm e}$ component 
(see Figure\,4), we apply the hydrogen density of 3.6\,cm$^{-3}$, obtained in the previous section. 
Similarly, we assume $n_{\rm H}$ = 3.9\,cm$^{-3}$ for the Ne\,{\footnotesize IX} resonance line 
that is dominated by the medium-$T_{\rm e}$ component. 
Using the mean elemental abundances obtained in Figure\,6,
and assuming $kT_i = kT_{\rm e}$ for both components, 
we obtain the plasma depth $L$ of $1.4 \times 10^{18}$\,cm and $5.9 \times 10^{18}$\,cm 
for the low-$T_{\rm e}$ (O\,{\footnotesize VII})  and medium-$T_{\rm e}$ (Ne\,{\footnotesize IX}) 
components, respectively. 
%% 追加20200707 -> Done. 
Therefore, if the resonance scattering is fully responsible for the anomaly in 
the $f$/$r$ ratio, the depth of the medium-$T_{\rm e}$ component must be 
about four times larger than that of the low-$T_{\rm e}$ component.
Both depth are $\lesssim$\,15\% of the SNR radius, reasonable for the region where 
the resonance scattering takes place.

%%順番変えた20200707 -> Footnote２文目追記
\begin{deluxetable*}{lccccc}
\tablecaption{Coefficients for the resonance scattering estimate. \label{tab:tau_rs}}
\tablewidth{0pt}
\tablehead{
\colhead{}& \colhead{$f$}	&	\colhead{$n_{\rm z}/n_{\rm Z}$}&	\colhead{$(n_{\rm Z}/n_{\rm H})_{\odot}$} &  \colhead{$A_Z$}&\colhead{$n_{\rm H}$}}
\startdata
O\,{\footnotesize VII}  $r$&	0.720&	0.13&	$4.90 \times 10^{-4}$ &0.35& 3.6\\ 
Ne\,{\footnotesize IX} $r$&	0.742&	0.29&	$8.71 \times 10^{-5}$ & 0.53& 3.9\\ 
\enddata
\tablecomments{$f$ is the oscillator strength. $A_Z$ is the abundance relative to the solar value. The unit of the last column is cm$^{-3}$.}
\end{deluxetable*}

\subsection{Charge Exchange}

\begin{figure*}
\gridline{\fig{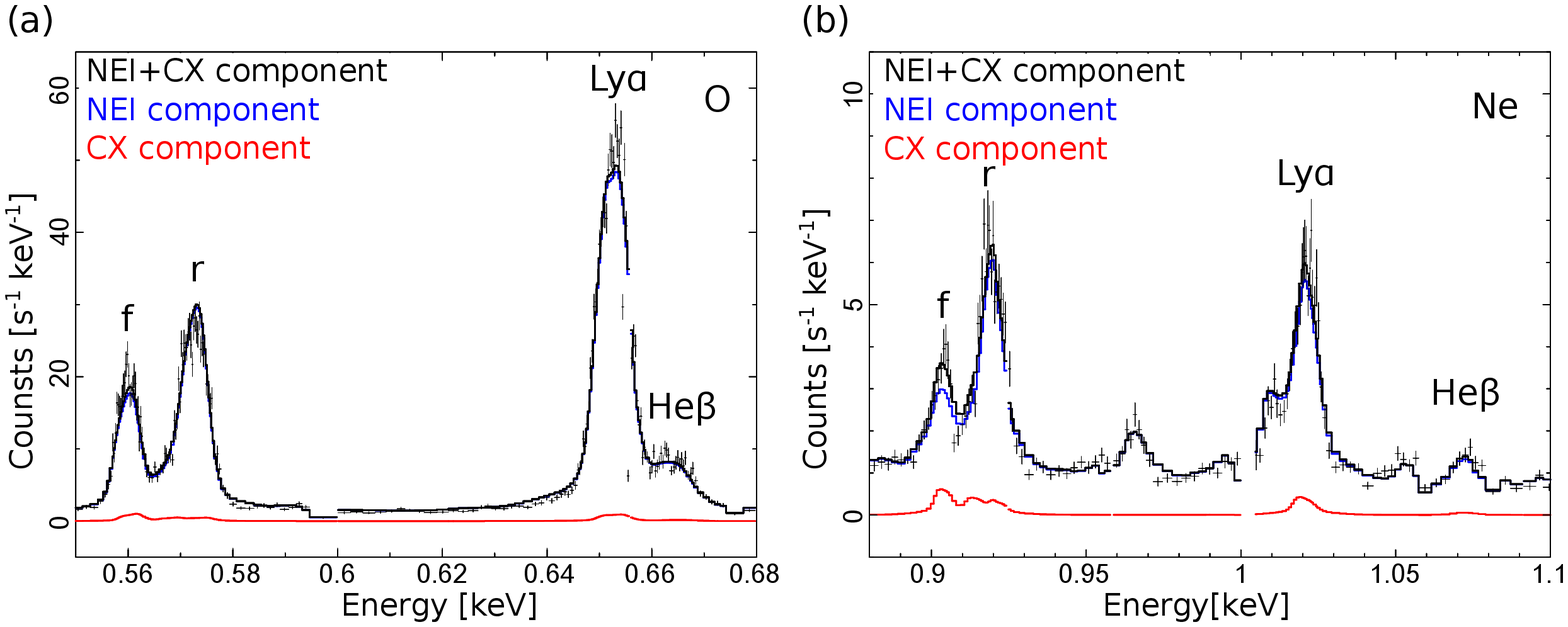}{0.8\textwidth}{}}
\caption{RGS1 spectrum around the O emission (a) and Ne emission (b), fitted with 
the model consisting of the three-temperature NEI (blue) and charge exchange (red) components. 
\label{fig:cx}}
\end{figure*}

Given the presence of the dense molecular clouds at the periphery of N132D 
\citep[e.g.,][]{Williams2006,Sano2015,Dopita2018}, charge exchange between 
highly-charged ions and neutral atoms or molecules would be a possible alternative 
that is responsible for the enhanced $f$/$r$ ratio.  In fact, evidence for charge exchange 
has been observed in several other SNRs that are interacting with dense clouds 
\citep[e.g., Pup A, Cygnus Loop:][]{Katsuda2012,Uchida2019}. 
We thus fit the full-band spectra from the deepest observation (Obs.\,ID: 0157161001) 
with the AtomDB-based charge exchange model \citep[{\tt vacx}:][]{Smith2012}, 
in addition to the three-component NEI model. Specifically, we introduce a {\tt vacx} 
model by assuming the ionization temperatures of O and Ne to be 0.24\,keV and 
0.39\,keV, respectively (based on Figure\,\ref{fig:ratio} a-2 and b-2). 
The O and Ne in this component are treated as free parameters. 
The parameter {\tt `model'} that determines the ($n$, $l$) distribution of the exchanged 
ions\footnote{$n$ and $l$ are the principal quantum number and orbital angular momentum, 
respectively. See http://www.atomdb.org/CX/acx\_manual.pdf for more details.}
is set to the default value `8', but this does not affect the results significantly. 
The addition of the charge exchange components slightly improves the C-stat value from 10426 to 
10417 (when all the parameters of the NEI components are fixed to the original best-fit values)  
or 10393 (when the parameters given in Table 3 are allowed to vary).
The close-up spectra around the O and Ne emission are shown in Figure\,\ref{fig:cx}, 
where the contribution of the charge exchange components is indicated by the red curves.
The deficit in the forbidden line flux in the NEI component is partially compensated by the charge exchange component.

The volume emission measure of the charge exchange component, defined as $\int{n_{\rm H} n_{\rm H^0} dV}$, is obtained 
to be $1.4 \times 10^{33}$\,cm$^{-3}$, where $n_{\rm H}$ is the density of hydrogen in the hot plasma, 
$n_{\rm H^0}$ the density of neutral hydrogen, and $V$ the volume where the charge exchange takes place.
Assuming $n_{\rm H}$ = 3.6--3.9\,cm$^{-3}$ (obtained in the previous section) and 
$n_{\rm H^0}$ = 100\,cm$^{-3}$ \citep{Dopita2018}, we estimate the volume of 
the charge exchange interaction region to be 3.6--3.8 $\times 10^{30}$\,cm$^{3}$. 
%%変更 20200707 -> Done.
Although charge exchange is expected to occur only in a narrow layer immediately behind 
the shock front \citep[up to a few percents of the SNR radius: e.g.,][]{Lallement2004}, 
the obtained volume is too small compared with the volume of the swept-up ISM 
($\sim 8.7 \times 10^{58}$\,cm$^3$). Therefore, if the $f$/$r$ ratio is enhanced solely 
by charge exchange, it is required for this process to take place only in local small regions, 
such as the south rim where the SNR forward shock is interacting with the molecular clouds \citep{Sano2015}.
The abundances of O and Ne are obtained to be $0.14_{-0.10}^{+0.58}$ and $1.4_{-0.49}^{+0.06}$, %誤差の範囲訂正 20200707
respectively. The latter is about three times higher than the Ne abundances of 
our best-fit NEI components as well as the LMC average. 
%%変更20200707 -> Done. 
%However, given the uncertainties in the currently available charge exchange models, we cannot rule out the possibility of the charge exchange contributing to the observed spectrum. 
This result, together with the small volume of the charge exchange interaction region, 
may suggest that the charge exchange scenario is less likely as the origin of the enhanced 
$f$/$r$ ratio, although the uncertainties in the currently available charge exchange models
are relatively large.

\subsection{Caveats and Future Prospects}

We have demonstrated that the high-resolution RGS spectra of N132D can reasonably be 
reproduced by the three-temperature plasmas in the NEI state plus the contributions 
of resonance scattering and/or emission induced by charge exchange 
(although the latter scenario requires an extremely small volume of the interaction region). 
However, it is not yet conclusive whether these effects are indeed at work, 
since our three-temperature model is only an approximation of the more realistic 
multiple-temperature plasma. 
The observations that enable spatially-resolved 
spectral analysis are crucial to provide conclusive evidence of these effects. 
If the resonance scattering indeed contributes to the observed spectrum, the resonance line flux 
is expected to be reduced most significantly at the region with the highest plasma column density. 
If the charge exchange is dominant, on the other hand, a spatial correlation between 
the forbidden-line enhancement and the ambient density is expected. 
We actually search for such evidence by extracting monochromatic images of 
O\,{\footnotesize VII} $r$ and $f$ emission from the first-order RGS data as was done 
by \cite{Uchida2019}, but no significant spatial variation in the $f/r$ ratio is found, 
owing to the insufficient spatial resolution and statistics. 
It should also be noted that the physics of charge exchange is still highly uncertain 
in both theoretical and observational aspects. 
Future improvement in the charge exchange models will help provide better constraints 
on the physics of SNR-ISM interaction with high-resolution spectroscopy.

\section{Conclusions} \label{sec:con}

We have presented high-resolution spectroscopy of N132D, the X-ray brightest 
SNR in the LMC, utilizing the XMM--Newton RGS data accumulated to date. 
The L-shell emission lines of Ar, Ca, and other elements have been detected from 
this SNR, for the first time. To enable quantitative analysis, we have generated 
spectral responses as carefully as possible, by taking into account the roll angle 
and wavelength dependences on the line spread function. 
The 0.3--2.0-keV spectrum can be well reproduced by the three-temperature 
optically-thin thermal plasmas in the NEI (ionizing) state, 
which reasonably approximates more realistic multiple-temperature plasmas. 
The measured elemental abundances are comparable to the average values of the LMC ISM, 
confirming that the soft X-rays from this SNR are dominated by the swept-up ambient medium.
The second-order RGS spectra, which we have analyzed for the first time, 
have successfully resolved the forbidden and resonance lines of 
Ne\,{\footnotesize IX}, allowing us to perform detailed plasma diagnostics 
using these lines. 
The observed forbidden-to-resonance flux ratios of O\,{\footnotesize VII} and 
Ne\,{\footnotesize IX} are slightly higher than expected for the typical NEI plasma. 
Either resonance scattering and/or charge exchange emission may account for this enhancement, 
although spatially-resolved high-resolution spectroscopy 
with future missions, like the {\it Athena} X-ray Integral Field Unit
\citep[X-IFU:][]{Barret2018}, is necessary to provide conclusive evidence. 

\acknowledgments

We thank Yoshitomo Maeda, Shinya Yamada, and Hikaru Suzuki for their helpful advice on data analysis and discussion. This work is partially supported by Grants-in-Aid for Scientific Research (KAKENHI) of the Japanese Society for the Promotion of Science (JSPS) grant Nos., 19H00704 (HY) and 19K03915 (HU).

\bibliography{ms_n132d_20200708}{}
\bibliographystyle{aasjournal}

\end{document}